%% file: main.tex
\documentclass{elsarticle}

\usepackage{graphicx}
\usepackage{tikz}
\usetikzlibrary{positioning}
\usetikzlibrary{shapes.geometric, arrows}
\usepackage{subfig}
\usepackage{amsmath}
\usepackage{amssymb}
\usepackage{color}
\usepackage{algorithm}
\usepackage{algcompatible}
\usepackage{comment}
\usepackage{hyperref}
\usepackage[noend]{algpseudocode}
\usepackage{wrapfig}
\algnewcommand{\LeftComment}[1]{\Statex \(\triangleright\) #1}

\newcommand{\skott}{{\sc skott}}

\newcommand*{\short}{\textcolor{red}}
\renewcommand{\vec}[1]{\mathbf{#1}}

\tikzstyle{retrieve} = [rectangle, rounded corners, minimum width=3cm, minimum height=1cm,text centered, draw=black, fill={rgb:red,4;blue,2;white,7}]

\tikzstyle{prep} = [diamond, minimum height = 3cm, text centered, draw=black, fill=green]
\tikzstyle{ad} = [rectangle, rounded corners, minimum width = 3cm, text centered, draw=black, fill={rgb:red,4;green,1;yellow,4}, node distance=3cm]
\tikzstyle{proc} = [rectangle, rounded corners, minimum width = 3cm, text centered, draw=black, fill={rgb:red,4;blue,2;white,7}, node distance=1cm]
\tikzstyle{dsp} = [rectangle, rounded corners, minimum width=2cm, text centered, draw=black,node distance=3cm, fill={rgb:red,4;blue,2;white,7}]
\tikzstyle{sub} = [rectangle, rounded corners, minimum width=2cm, text centered, draw=black,node distance=1cm]
\tikzstyle{rtb} = [rectangle, rounded corners, minimum width=2cm, text centered, draw=black, node distance=1.2cm, fill={rgb:red,4;green,1;yellow,4}]
\tikzstyle{anc} = [circle, radius=2pt]
\tikzstyle{arrow} = [thick,->,>=stealth]

\begin{document}

\begin{frontmatter}
\title{A New Optimization Layer for Real-Time Bidding Advertising Campaigns}

\author{Gianluca~Micchi, Saeid~Soheily-Khah, Jacob~Turner}
\cortext[mycorrespondingauthor]{Corresponding author}
\ead{research.paris@skylads.com}
\address{Skylads, 75008 Paris, France}
\date{}

\begin{abstract}
While it is relatively easy to start an online advertising campaign, obtaining a high Key Performance Indicator (KPI) can be challenging.
A large body of work on this subject has already been performed and platforms known as DSPs are available on the market that deal with such an optimization.
From the advertiser's point of view, each DSP is a different black box, with its pros and cons, that needs to be configured.
In order to take advantage of the pros of every DSP, advertisers are well-advised to use a combination of them when setting up their campaigns.
In this paper, we propose an algorithm for advertisers to add an optimization layer on top of DSPs.
The algorithm we introduce, called \skott, maximizes the chosen KPI by optimally configuring the DSPs and putting them in competition with each other.
\skott\ is a highly specialized iterative algorithm loosely based on gradient descent that is made up of three independent sub-routines, each dealing with a different problem: partitioning the budget, setting the desired average bid, and preventing under-delivery.
In particular, one of the novelties of our approach lies in our taking the perspective of the advertisers rather than the DSPs.
Synthetic market data is used to evaluate the efficiency of \skott\ against other state-of-the-art approaches adapted from similar problems.
The results illustrate the benefits of our proposals, which greatly outperforms the other methods.
\end{abstract}
\begin{keyword}
Keywords: Demand Side Platform (DSP), Online Advertising, Gradient Descent, Optimization, Real Time Bidding (RTB)
\end{keyword}

\end{frontmatter}

\input{introduction}

\input{relatedWork}

\input{problem}
\input{budget}
\input{bid}
\input{pacing}

\input{experiments}

\input{conclusion}

%
\appendix
\input{market}
\input{periodicity}
\input{dataProcessing}
\input{algorithms}

%
\bibliographystyle{plain}
\bibliography{skylads}

%

\end{document}

%% file: introduction.tex
\section{Introduction}
\label{sec:introduction}

Online advertising is a vast market, worth several billion USD per year \cite{IAB2017}.
It is easy to understand the importance of optimization in such a market: every percent of increase in efficiency has a value in the order of millions of USD.
From the advertiser's point of view, however, optimization is often very difficult due to the constraints imposed by the structure of the ecosystem of online advertising.
Let us see why.

There are currently two main paradigms in the market: sponsored search and Real-Time Bidding (RTB) auctions.

Sponsored search, the main source of revenues for search engines, consists in showing relevant ads whenever a user inputs a query.
For example, typing ``football shoes" in the search bar will provide the user with several links to online shops selling sports apparel.
In order to appear amongst the sponsored links, an advertiser must place a bid on the queries or keywords to which it wants to be connected.
Typically, advertisers pay if and only if their sponsored link is clicked on;
this creates a collaborative effort between search engine and advertiser to show the most promising ad.
Sponsored search has been the subject of lots of research papers over the years, dealing with, for example, budget optimization \cite{Feldman2006, Karande2013} and click-through rate prediction \cite{Richardson2007,Zhang2014e}.

The RTB paradigm, which is the focus of our optimization work, is inherently different.
In this paradigm, advertisers participate in auctions to buy the available space on a website.
The winning advertiser is allowed to display its ad (in the technical jargon, it has bought an ``impression'').
Unlike sponsored search, advertisers pay for all impressions they buy, even if they don't ``generate a click'', i.e., users don't click on them to be redirected to the advertiser's page.
This changes everything as it removes any interest for the auctioneer to find an ad that is a good match to the current inventory, a task that is now completely left to the advertisers.
As a consequence of these differences, the optimization results obtained in sponsored search are not directly applicable to RTB campaigns.

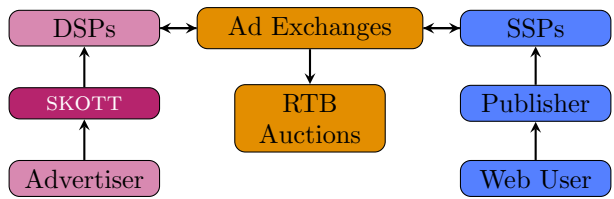
\begin{figure}[!b]
\centering
\begin{tikzpicture}
\node [ad] (ad) {Ad Exchanges};
\node [dsp, left of=ad] (dsp) {DSPs};
\node [dsp, right of=ad, fill={rgb:blue,3;cyan,1;white,2}] (ssp) {SSPs};
\node [sub, below of=dsp, text=white, fill={rgb:red,4;blue,2;white,1}] (skott) {\skott};
\node [sub, below of=skott, fill={rgb:red,4;blue,2;white,7}] (adv) {Advertiser};
\node [rtb, below of=ad, align=center] (rtb) {RTB \\ Auctions};
\node [sub, below of=ssp,fill={rgb:blue,3;cyan,1;white,2}] (pub) {Publisher};
\node [sub, below of=pub, fill={rgb:blue,3;cyan,1;white,2}] (user) {Web User};

\draw [thick, ->, >= stealth] (ad) -- (dsp);
\draw [thick, ->, >= stealth] (dsp) -- (ad);
\draw [thick, ->, >= stealth] (ad) -- (ssp);
\draw [thick, ->, >= stealth] (ssp) -- (ad);
\draw [thick, ->, >=stealth] (skott) -- (dsp);
\draw [thick, ->, >=stealth] (adv) -- (skott);
\draw [thick, ->, >=stealth] (ad) -- (rtb);
\draw [thick, ->, >= stealth] (pub) -- (ssp);
\draw [thick, ->, >= stealth] (user) -- (pub);
\end{tikzpicture}
\caption{A simple, approximate representation of the RTB auctions market structure. The proposed algorithm, \skott, would work as an interface between advertisers and DSPs.}
    \label{fig:market}
\end{figure}

As we already mentioned, this optimization process is made difficult by the very structure of the market that, in its simplest approximation, looks like this (cf. Fig.~\ref{fig:market}):
The central entity is the AdExchange, whose job is to run real-time bidding auctions and assign every available inventory (i.e., the screen space on which the advertising should be published) to its corresponding winning advertiser;
On one side of AdExchanges there are Supply Side Platforms (SSPs), which provide the inventory and are in direct contact with the publishers (e.g., the owner of a web page);
On the other side, the one that interests us the most, there are Demand Side Platforms (DSPs): they bid on the available inventory on behalf of the advertisers according to the advertisers' necessities.
Each step of this chain brings its own constraints and optimizations: for example, a DSP might work only with certain AdExchanges, effectively limiting the amount of inventory available to the advertiser, but it also might offer better performances on some particular indicators due to internal optimization algorithms.

It is easy for an advertiser to set up a campaign with a DSP and monitor its effectiveness by calculating certain Key Performance Indicators (KPIs).
On the flip side, however, the use of a DSP prevents the advertiser from directly determining the bid on each individual auction, only allowing it to fix some average parameters to associate with larger sets of impressions.
We call each of these abstract entities made of a set of impressions and its corresponding parameters a \emph{media object}.
Each media object can almost be treated as a separate entity with its own associated budget that can change over time, the only global constraint being that the total budget of the campaign is fixed.
A single media object makes the optimization easy to handle but inefficient because it treats all impressions in the same way, bidding roughly the same amount of money for all impressions and showing the same advertising to all users.
A correct choice of the parameters of the media objects can, therefore, have a huge impact on the global optimization of the campaign.

But this setup leaves advertisers with many questions:
What is the most appropriate DSP to use amongst the many available on the market?
How to parametrize it?
How much budget to assign to each of its media objects?
These questions are often answered by human experts that base their decisions on past experience, intuition, and some off-line data analysis.
However, there is no guarantee that the goals set by the experts are reachable, nor that they are optimal.

In recent decades, researchers have focused mainly on market models \cite{Conchar2005,Gusev2014,Leeflang2000,Pietersz2010,Zhang2010} 
and bidding algorithms for DSPs \cite{Chen2011,Zhang2014,Donnellan2015,Grigas2017,Liu2017,Ren2018}.
However, to the best of our knowledge, no paper tackles the optimal management of a campaign from the point of view of an advertiser that uses a DSP.
The new constraints that come with such a perspective make other optimization algorithms studied in the literature difficult to compare with ours.
For example, our need to partition the budget arises from the necessity to work with media objects: when an impression arrives that is a good fit to a particular media object, we need to be sure that the media object has a sufficient budget to spend.
This problem does not exist when one can decide how much money to spend on a single impression basis, with the only constraint of the total campaign budget.

The main contribution of this paper is the \skott\ algorithm, that solves this understudied problem:
\skott\ automatically handles advertising campaigns, finding the best parameters to put inside each DSP in order to maximize the performance;
it also allows the contemporary use of different DSPs that are put in competition to further increase the performance.
Therefore, not only it gives a recipe to fully take advantage of any single DSP, but it also adds a new layer of optimization on top of them.
The algorithm reacts quickly to market variations and scales linearly with respect to the number of media objects.

The  remainder  of  this  paper  is  organized  as  follows. 
In Section~\ref{sec:related_work} we discuss a few algorithms that deal with a similar problem and have been an inspiration for our work.
In Section \ref{sec:problem} we state the problem at hand and our goals. 
Sections \ref{sec:budget}, \ref{sec:bid}, and \ref{sec:pacing} deal with the three independent sub-routines that compose \skott.
The conducted experiments and results obtained are discussed in Section \ref{sec:experimental_results}.
Conclusions are drawn in Section~\ref{sec:conclusion}.
Finally, appendices give details on the creation of the synthetic market data, the algorithms we chose for comparison of the results, and the technical part of the implementation.

%% file: relatedWork.tex
\section{Related work}
\label{sec:related_work}

The problem of how to spend a budget in order to maximize the profit during an advertising campaign has never been studied, to the best of our knowledge, from the point of view of an advertiser using DSPs.
Nevertheless, there are many works in the literature that are relevant because they try to solve a similar problem.
We will cite and discuss some of them in this section, mainly to highlight what is the difference in our approach.

In \cite{Feldman2006} the authors propose to use a randomized uniform strategy for choosing how much to bid on every keyword in a sponsored search. 
That could be applied to our problem of Real-Time Bidding auctions through the use of DSPs problem by analogy, associating every keyword with a different media object.
(We refer to the introduction and to the problem statement for the definition of media objects.)
However, their model assumes a complete knowledge of the bidding landscape, that is, the probability distribution of the winning bids for each impression.
This is information that advertisers don't have in the case of RTB auctions through DSPs.
Also, the model in \cite{Feldman2006} requires the bidding landscape to be static, a hypothesis that we don't require.

The problem of collecting the largest possible reward of an advertising campaign with a constraint on the budget can be also written in the linear programming formalism.
This is done, for example, in \cite{Chen2011}. 
There, however, the authors: 1. take the point of view of a DSP, and 2. try to optimize the revenue of publishers and not of advertisers.
In particular, they assign each piece of inventory to a different advertising campaign assuming that the total Cost Per Click (CPC) of each campaign is fixed.
This last point is clearly not the case when we look at it from the advertiser's side because, as we said, we consider the case where advertisers pay for the inventory they buy regardless of it being clicked on.
This implies that the CPC is then determined by the ratio between the average cost per impression and the click-through ratio of a media object, therefore it's not constant, as we will thoroughly see when we analyze the \skott\ algorithm.

A similar work, explicitly based on \cite{Chen2011}, is \cite{Liu2017}, where a linear programming algorithm is proposed for DSPs wanting to maximize their revenues.
Besides the different perspective, the authors assume a fixed Click-Through Rate (CTR), known in advance.
This is not required in our approach, where we infer the CTR from market data in real time and the only assumption we make on its analytical form is that it varies slowly with the bid.
A linear programming approach inspired by these works will be tested against \skott\ in the simulations.

A third approach to the problem is due to reinforcement learning.
In this case, the bidder is considered as an agent that learns how much to bid for every individual auction.
To direct its choice, it is aware of the budget constraints, the goals of the campaign, and all context information from the impression it is trying to buy.
This approach is studied for example in \cite{Cai2017} where, due to the huge size of the space of possible actions to take, the authors help the decision process by using a model-based approach.
This approach is extremely interesting but ineffective in our case for two main reasons: first of all, the constraints under which we work prevent us from accessing individual auctions; and, secondly, we obtain information not in real time but in batches that come in at larger time scales (roughly an hour).
Nevertheless, the reinforcement learning approach is tested against \skott\ in the simulations, adapting it to allocate the budget to the different media objects according to their results.

%% file: problem.tex
\section{Problem statement}
\label{sec:problem}

As stated in the introduction, advertisers can participate in RTB auctions by using a DSP.
Typically, several DSPs are employed in order to increase the amount of people reached and better respond to business necessities.
Each DSP needs to be configured.
The details of the configuration might differ from DSP to DSP, but there is a central core of abstraction that is common to all of them.
We call it a \emph{media object}: it is a set of instructions given by the advertisers, some of which are qualitative and set once and for all at instantiation while others are quantitative and can be changed at any moment.
An example of the former is the creative associated to the media object, i.e., the actual advertising being shown.
Another example of qualitative instructions are the filters on incoming auctions that select on which impressions to place a bid depending on the user and inventory characteristics.
The hourly budget and the base bid for the auctions, instead, belong to the latter.
It is important to note that the media object is the most precise layer of abstraction that is accessible to advertisers.
The only influence that they can have on the auctions, and therefore their only possibility for optimization, lies in the parameters of the media objects. 


We consider an advertising campaign to be defined by: a total budget $\mathcal B$, a start date, an end date, a desired spend profile (i.e., the amount of total money spent at any moment during the campaign), and a collection of $K$ media objects spread across different DSPs.
The typical duration for a campaign is of the order of a few weeks up to a few months, while the value of $K$ depends heavily on the campaign and can vary from as few as 1 to over 100\,000. 

During the campaign lifetime, the advertiser receives information on the behavior of every media object from the different DSPs. Each data point contains hourly information on the impressions bought, the clicks generated, the money spent, and possibly other such quantities.
Since advertisers don't have access to the auctions individually but only through the media objects, they should only consider the average effect of their optimizations over all impressions.
Therefore, for each media object we take an hourly average of the information received.
In practice, we consider only the Click-Through Rate (CTR, the ratio of clicks generated to impressions bought) and Cost-Per-Click (CPC, the ratio of money spent to clicks generated).

The main goal of our algorithm is to change the media objects parameters in such a way as to optimize a certain KPI while keeping the desired delivery over time.
In order to demonstrate a practical case, we have chosen to optimize the total number of clicks generated in the campaign.
This is often a valid indicator to optimize because a user that is interested to purchase something from the advertiser's website will probably click directly on the ads, while only a small fraction of the people that click on the ads will actually make a purchase.
The click is then correlated to the monetary return of the campaign while not being as rare as an actual purchase.

%% file: budget.tex
\section{\skott: Budget partitioning}
\label{sec:budget}

\skott\ is an iterative algorithm made up of three subroutines: 
budget partitioning, that rewards high-quality media objects by giving them more money;
base bid setting, that controls the bid of each media object separately with the goal of increasing the media object's quality;
and pacing control, that prevents under-delivery.
Figure~\ref{fig:algo} provides a schematic view of the different steps of the algorithm.
The three sub-routines act independently one from another and can therefore be analyzed in any order.
In this section we deal with the budget partitioning that defines which percentage of the hourly budget should be allocated to each media object.

\begin{figure}[!ht]
 \centering
 \begin{tikzpicture}
 \node (start) [retrieve, align=center] {\small Data retrieval \\ and pre-processing};
 
 \draw [right of=start, fill={rgb:red,4;blue,3;white,1}, rounded corners, xshift=1cm, yshift=-1.6cm] (0,0) rectangle (3.85,3.5);
 \node [text=white, right of=start, xshift=2.8cm, yshift=1.5cm] {\skott};
 
 \node (bid) [proc, right of=start, xshift=3cm] {\small Base bid setting};
 \node (budget) [proc, above of=bid] {\small Budget partitioning};
 \node (pacer) [proc, below of=bid] {\small Pacing control};
 
 \draw [right of=start, xshift=.85cm, thick] (0,-.5) -- (0,.5);
 \draw [right of=start, xshift=.85cm, thick] (-.25,0) -- (.25,0);
 \node (anc3) [right of=start, xshift=.85cm]{};
 \draw [thick] (start) -- (anc3);
 \draw [arrow] (anc3) |- (budget);
 \draw [arrow] (anc3) -- (bid);
 \draw [arrow] (anc3) |- (pacer);
 \node (dsp) [dsp, above of=anc3, xshift=1.5cm, yshift=-.5cm] {\small DSP};
 \draw [arrow] (dsp) -| (start);
 
 \node [circle, draw=none, left of=dsp, xshift=-1.1cm, yshift=.25cm] {market data};
 \node [circle, draw=none, right of=dsp, xshift=1.4cm, yshift=.25cm] {new instructions};
 
 \draw [xshift=6cm, thick] (0,-.5) -- (0,.5);
 \draw [xshift=6cm, thick] (-.25, 0) -- (.25,0);
 \node (anc) [anc, right of=bid, xshift=1cm] {};
 \node (anc2) [anc, right of=anc] {};1
 \draw  [thick] (bid) -- (anc);
 \draw [thick] (anc) -- (anc2);
 \draw [arrow] (anc2) |- (dsp);
 \draw [thick] (budget) -| (anc);
 \draw [thick] (pacer) -| (anc);
 \draw [thick, xshift=7cm]  (0,0) -- (0,.5);
 \draw [thick, xshift=7cm]  (0,0) -- (-.5,0);
 
  \end{tikzpicture}
  \vspace{0.3cm}

  \hrule
  \skott\ algorithm\\
  \hrule
  \begin{algorithmic}[1]
  \Repeat { at every hour}
  \Procedure{Retrieve and preprocess data}{\ref{sec:data_pre-processing}}
  \EndProcedure
  \Procedure{Budget partitioning}{section \ref{sec:budget}}
  \EndProcedure
  \Procedure{Base bid setting}{section \ref{sec:bid}}
  \EndProcedure
  \Procedure{Pacing control}{section \ref{sec:pacing}}
  \EndProcedure
  \Procedure{Finalize}{}  
  \State send new parameters to the media objects
  \EndProcedure
  \Until {the campaign is finished}
  \end{algorithmic}
  \hrule
  \caption{A schematic view of the \skott\ algorithm.}\label{fig:algo}\hrule
\end{figure}

A budget partition is a vector of $K$ weights $\vec w$.
Each element $w_i$ represents which fraction of the total budget is assigned to the corresponding media object $i$.
A uniform distribution, where all media objects are assigned the same budget, is represented by the vector $\vec u = \left(k^{-1}, k^{-1}, ..., k^{-1}\right)$.
A greedy distribution is when a single media object takes all the available budget and is represented by the vector $\vec w_g^{(i)} = \left(0, ..., 1, ..., 0\right)$.

The ideal algorithm for budget partitioning should:
\begin{itemize}
\item return a list of non-negative weights that sum to one at every decision epoch $t$,
\item optimize a specific KPI (in our case the total number of clicks),
\item promptly react to changes in the market, be they sudden or slow.
\end{itemize}

A very important point to consider when devising the algorithm is that the data must be bought through winning auctions.
Reducing the budget of a media object well below the expected CPC will result in no clicks being bought, thereby gaining little to no useful information to estimate the quality of the media object.
An advertiser may spend some time and money to explore the market randomly, then concentrate their money on the best performing media objects.
This would probably lead, in average, to an increase in the return of the campaign; but the price to pay is a high risk to get stuck on a sub-optimal media object.
A more dynamic algorithm that keeps exploring over time seems therefore a more reasonable choice.

There is a balance to strike between exploration and exploitation.
The former is expensive, but mitigates risks and gives a better long-term investment.
The latter increases the short-term reward, but might prove catastrophic over the long-term, locking the investment on media objects that are ultimately bound to fail. 

\begin{figure}[!t]
\centering 
\hrule
\vspace{1pt}
Algorithm for budget partitioning
\hrule
\begin{algorithmic}[1]
\Function{\textsc{budget partitioning}}{}
\LeftComment{$t$: epoch index}
\LeftComment{$\vec C_t$: clicks obtained by the media objects during $t$}
\LeftComment{$\vec B_t$: budgets assigned to the media objects before $t$}
\LeftComment{$\vec{\hat C}_{t-1}$: cumulative discounted clicks before $t$ (unless $t=0$)}
\LeftComment{$\vec{\hat B}_{t-1}$: cumulative discounted budgets before $t$ (unless $t=0$)}
\LeftComment{$\vec w_t$: previous repartition of the budget}
\LeftComment{$\vec u$: uniform repartition of the budgets over the media objects}
\LeftComment{$\lambda_t$: regularization parameter}
\LeftComment{$\alpha$: learning rate}
\LeftComment{$\gamma$: discount factor}

\Procedure{\textsc{update the discounted quantities}}{}
\If {$t=0$}
  \State  $\vec{\hat C}_0 \gets \vec C_0$ \nonumber
  \State  $\vec{\hat B}_0 \gets \vec B_0$ \nonumber
\Else
  \State  $\vec{\hat C}_t \gets \gamma \ \vec{\hat C}_{t-1} + \vec C_t$ \nonumber
  \State  $\vec{\hat B}_t \gets \gamma \ \vec{\hat B}_{t-1} + \vec B_t$ \nonumber
\EndIf
\EndProcedure

\Procedure{\textsc{calculate gradient of loss function}}{}
\State  $\vec Q_t \gets \vec{\hat C}_t/\vec{\hat B}_t$ \nonumber
\State  $\vec{\tilde Q}_t \gets \vec Q_t / \max(\vec Q_t)$ \nonumber
\State  $\nabla \mathcal L_t \gets - \vec{\tilde Q}_t + \lambda_t \cdot (\vec w_t - \vec u)$ 
\EndProcedure

\Procedure{\textsc{update weights}}{}
\State $\vec c  \gets \exp\left( - \alpha \cdot \nabla \mathcal L_t\right)$
\State $\vec w_{t+1} \gets \vec w_t * \vec c / \sum_i w_{i,t} c_i$ 
\Comment{$*$: element-wise multiplication}
\EndProcedure

\State {\bf{return}} $\vec w_{t+1}$
\EndFunction
\end{algorithmic}
\hrule
\caption{Algorithm for budget partitioning}\label{alg:budget}
\hrule
\end{figure}

\subsection{The update rule}
The algorithm we propose is a variation of the exponentiated gradient descent method originally proposed in \cite{Kivinen1997}.
At each iteration, it updates the weight vector $\vec w$ in order to optimize the KPI.
The algorithm is resumed in Figure~\ref{alg:budget}.
It makes use of the concept of a reward assigned to each media object at every decision epoch. 
The reward is a numerical way to estimate how well the media object did in the epoch.
We will see explicitly what it looks like later on.

Given a vector of weights at epoch $t$, $\vec w_t = [w_{1,t}, w_{2,t}, ..., w_{K,t}]$, describing the distribution of the budget allocated to each of the $K$ media objects during the $t$-th hour of the campaign, the algorithm will return a new vector of weights $\vec w_{t+1}$ that is closer to the minimum of the loss function
\begin{equation}
\label{eq:loss}
\mathcal L_t(\vec w_t) = - \sum_i R_i(\vec w_t) + \frac{\lambda_t}{2} \lVert \vec w_t - \vec u \rVert ^2 .
\end{equation}
Here, $R_i$ is the reward associated to every media object,
$\lambda > 0$ is the regularization parameter (that can depend on the epoch $t$),
and $\vec u$ is the vector of uniform distribution with all entries equal to $1/K$ that we introduced before.
The effect of the first term of the loss function is to favor the repartitions giving larger reward.
The second term, known as the regularization term, requires the repartition $\vec w$ to be close to the uniform distribution $\vec u$.
In other terms, it enforces the exploration of the market, with the consequences discussed in section~\ref{sec:problem}.
The relative importance of the exploration is therefore given by the numerical parameter $\lambda$ that can be set at will.
The easy interpretation of this parameter and its conceptual relevance is an important feature of our algorithm.
We will discuss how to choose it at the end of this section.

The update rule defined by the exponentiated gradient descent is the following:
\begin{equation}
\vec w_{t+1} = \frac{\vec w_t \cdot \exp\left( - \alpha * \nabla \mathcal L_t(\vec w_t) \right)}{\displaystyle \sum_i [w_{i, t} \cdot \exp\left( - \alpha * \nabla_i \mathcal L_t(\vec w_t) \right)]},
\label{eq:updateRule}
\end{equation}
where $\alpha$ is a real positive parameter known as the \emph{learning rate}
and $\nabla$ indicates the gradient of a function with respect to the vector of weights $\vec w$.

\subsection{Explicit calculation of the derivative of the loss function}
The rest of this section is devoted to write what is the value of $\nabla \mathcal L_t(\vec w_t)$ that is needed to update the weights.
To do so, we need to explicitly define what is the reward and to find its gradient.
In general, the reward is given by the goal of the advertising campaign.
As we already mentioned, we will use the maximization of the number of clicks as an example.
The reward is then simply the number of clicks that a media object obtains during an epoch: $\vec R = \vec C$.
Its gradient represents the relative change in the number of clicks that a media object would have generated if we had given it a slightly different budget.
This clearly can not be obtained directly from the market.
Our solution is to model the relation between clicks and budget analytically, derive the gradient, and then approximate it using the sampled results from the market. 
In equations, this reads:
\begin{gather}
\label{eq:reward}
R_{i,t}(\vec w_t) = Q_{i,t} (w_{i, t}) \cdot w_{i,t}; \\
\label{eq:reward_der}
\nabla \sum R_{i, t} = \vec Q_t (\vec{w_t}),
\end{gather}
where $\vec Q_t (\vec{w_t})$ is the (unknown) vector of the coefficients that represents conceptually the quality of the media objects.
Notice that to pass from Eq.~\ref{eq:reward} to Eq.~\ref{eq:reward_der} we have made the assumption that $\vec Q_t (\vec{w_t})$ varies slowly with the weight vector so that its derivative becomes negligible.
This is clearly an approximation, but a useful one whose price we happily pay.

Since $\vec w_t = \vec B_t / \sum_i B_{i,t}$, where $\vec B_t$ is the vector of budgets associated to each media object at time $t$, the quality vector can be written as:
\begin{equation}\label{eq:no_discount}
\vec Q_t = \frac{\vec R_t}{\vec w_t} = \sum_i B_{i,t} \cdot \frac{\vec C_t}{\vec B_t}.
\end{equation}
Let us notice that $\vec Q$ is quite similar to the vector of inverse CPCs, the two differences being the (unimportant) global positive multiplicative factor $\sum_i B_{i,t}$ and the presence at the denominator of $\vec B_t$ instead of $\vec S_t$ (the budget allocated instead of the money actually spent during the epoch).
This is in accordance with our intuitive identification of $\vec Q$ with the quality of a media object because lower CPCs are desirable.

Let us also mention that, since the quality of the media objects depends on external factors, a rescaling is needed to ensure the relative importance of the regularization parameter (hence the uselessness of the global multiplicative factor $\sum_i B_{i,t}$).
We thus use the rescaled quality $\vec{\tilde Q}_t$ defined as:
\begin{equation}
\vec{\tilde Q}_t = \frac{\vec Q_t}{\text{max}(\vec Q_t)},
\end{equation}
which ensures all the elements of the vector to be positive and not larger than 1.

Under these conditions, we can rewrite the derivative of Equation~\ref{eq:loss} as:
\begin{equation}
\nabla \mathcal L_t(\vec w_t) = - \vec{\tilde Q}_t + \lambda_t (\vec w_t - \vec u).
\end{equation}

\subsection{Fighting the noise}
Due to the stochastic nature of the data coming from the market, there are a few corrections to make to the model of the quality vector in order to improve the precision and the stability of the results.

Let us consider the quality factor as defined in Eq.~\ref{eq:no_discount}.
The problem is that clicks are extremely rare:
A typical CTR is 0.1\%, meaning that only one impression out of a thousand generates a click.
However, it is always possible, albeit rare, that a media object buys a small amount of impressions and generates a click.
This is, of course, just sampling noise due to the very nature of the quantities we are dealing with.
However, if not taken into account, it would dominate the response of the algorithm and lead to very unstable situations.
Even worse, it could lock the algorithm to put all its money into a single, sub-optimal media object for a long time. 
To deal with that we make two corrections.

First of all, we put a hard bound on the gradient between $-10/\alpha$ and $+10/\alpha$, $\alpha$ being the learning rate of the gradient descent, to avoid exploding exponentials.
This is very simple and straight-forward, but it successfully prevents media objects with unusually large rewards to take all the budget.

Then, we claim that a better estimation of the value of the quality of a strategy can be done using a cumulative discounted version of the clicks and budgets, i.e., a variable that takes into account not only the latest data but also past data weighted by a discount factor $\gamma$:
\begin{equation}
\label{eq:quality}
\vec Q_t = \mathcal B_t \cdot \frac{\vec{\hat C}_t}{\vec{\hat B}_t},
\qquad
\vec{\hat C}_t = \vec C_t + \gamma \ \vec{\hat C}_{t-1},
\qquad
\vec{\hat B}_t = \vec B_t + \gamma \ \vec{\hat B}_{t-1},
\end{equation}
where we call $\vec {\hat C}$ the vector of cumulative discounted clicks initialized with the rule $\vec{\hat C}_0 = \vec C_0$ (and similarly for the vector of cumulative discounted budgets $\vec {\hat B}$). 
Here, $\gamma \in [0, 1]$ controls the importance of past data in the estimation of the quality of the media object:
When $\gamma = 0$ we have no memory and $\vec{\hat C}_t = \vec C_t$ (same for the budgets); we are back in the situation represented by (\ref{eq:no_discount}).
On the other hand, $\gamma = 1$ implies that the data collected at time $t_0$ is considered relevant for all $t > t_0$ and is never to be forgotten.
This is desirable only when the quality is guaranteed to be constant.
Since this is not the case, we use a $\gamma < 1$ to slowly forget data that is no longer relevant.
We can fix the exact value of $\gamma$ by choosing a time scale for our campaign.
If, for example, we want to forget data that is a week old, we can put $n_d = 7$ and solve the equation $\gamma^{n_d} = 1/e$ which yields $\gamma = e^{-1/n_d} \approx 0.87$.
A first order approximation to this result can be obtained with $\gamma \approx 1 - 1/n_d \approx 0.86$.

\subsection{The regularization parameter}
We have said that an important feature of our algorithm is the relevance of the regularization parameter, that decides the trade-off between exploration and exploitation.
Here, we explain what we chose in our simulations and why.

The regularization parameter that we used is 
\begin{equation}
\lambda_t = \eta \cdot K \cdot \gamma_r^{d(t)},
\end{equation}
where $\eta$ is a positive number that determines the exploration-exploitation trade-off (in our simulations it is set to 1),
$K$ is the number of media objects,
$\gamma_r$ is (another) discount factor that determines when should exploitation dominate over exploration,  
and $d(t)$ is the number of days that have passed since the beginning of the campaign.

The interest in rescaling with $K$ comes from the advantage of keeping the product $\lambda_t * \vec u$ that appears in the gradient of the loss function independent from the number of media objects.
(Remember that $\vec u$ is the uniform distribution, whose elements are all $k^{-1}$.)
This grants a comparable greediness (measured for example as the KL-divergence from the uniform distribution, see Equation~\ref{eq:kl}) when running on campaigns with vastly different number of media objects.

The presence of the term $\gamma_r^{d(t)}$ stems instead from the advantage of having a larger exploration at the beginning of the campaign and a larger exploitation towards the end, where we want to monetize the knowledge we have acquired.
The numerical value of $\gamma_r$ is determined in the same way as for the discount factor in the quality vector, just using a different time-scale.
If, for example, we want to keep a large exploration for 20 out of the 30 days of the advertising campaign, we would fix $\gamma_r = 1 - 1/20 = 0.95$.

%% file: bid.tex
\section{\skott: Base bid setting}
\label{sec:bid}

In this section, we present the algorithm that dynamically changes the base bid of a media object.

The base bid of a media object represents a sort of \emph{default value} that is adjusted by the DSP depending on how valuable it deems a certain piece of inventory for said media object. 
Many DSPs make this adjustment by multiplying the base bid by a score calculated from data about a specific item of inventory.
Typically, however, the base bid will represent the average bid offered during the campaign.

Clearly, a high base bid will lead to chronic overbidding.
This is indicated by the fact that the average cost per impression is significantly below the bid, assuming that the inventory is priced based on the second highest bid (as is overwhelmingly the case, cf. \cite{yuan2014survey}). 
Overbidding is very risky because it might lead to a very large expense on non-valuable impressions if another player in the market is making the same mistake.
Conversely, a low base bid can cause the inability of a media object to buy inventory deemed valuable by the DSP, leading to an under-utilized budget.

Still following the example for which our goal is to maximize the number of clicks, we write $\mathcal L_t = - \vec{C_t}$.
Notice that we don't put any regularization parameter because we don't put any constraints on the base bids, so far.
Now, we need to express the number of clicks as a function of the bids and then use gradient descent to maximize the function.
In practice, the function that we use is:
\begin{equation}
\label{eq:loss_bid}
\mathcal L_t(\vec b_t) = - \vec{C_t}(\vec b_t) = - \frac{\vec S_t(\vec b_t)}{\vec{CPC}_t(\vec b_t)}\\,
\end{equation}
where $\vec S_t (\vec b_t)$, and $\vec{CPC}_t (\vec b_t)$ are, respectively, the total amount of money spent and the resulting Cost Per Click in the previous epoch as a function of the base bid, each element of the vectors representing a different media object.
In line of principle, the loss function should maximize the amount of money spent while decreasing the CPC of each inventory piece.
These two objectives are contrasting: to spend more money one should increase the base bid to have access to more inventory, but to buy cheap clicks one should reduce the base bid.

In the following, we analyze separately the functions that appear on the right hand side of (\ref{eq:loss_bid}), starting with the CPC.
For sake of simplicity, since all the bids act independently, we will just give the solution using scalar quantities.
The full result is presented at the end of the section and is also resumed in Figure~\ref{alg:bid}.

\begin{figure}[!h]
\hrule
\centering 
\vspace{1pt}
Algorithm for setting the base bid
\hrule
\begin{algorithmic}[1]
\Function{\textsc{base bid setting}}{}
\LeftComment{$N_t$: number of impressions bought}
\LeftComment{$B_t$: allocated budget}
\LeftComment{$S_t$: money actually spent}
\LeftComment{$C_t$: clicks obtained}
\LeftComment{$b_t$: base bid}
\LeftComment{$\beta$: median winning bid over all the inventory}
\LeftComment{$\tau$: under-delivery threshold}
\LeftComment{$\alpha$: learning rate} 
\LeftComment{$u_b$: upper bound on the bid} 
\LeftComment{$l_b$: lower bound on the bid} 
\LeftComment{$B_\text{min}$: minimal tolerance for zero budget}

\Procedure{\textsc{calculate gradient of loss function}}{}

\If{$B_t < B_\text{min}$}
\State $\nabla \mathcal L(b_t) \gets 0$
\Comment{If no budget is set, ignore the media object.}
\ElsIf{$S_t=0$}
\State $\nabla \mathcal L(b_t) \gets - 1/\alpha$ 
\Comment{If no money is spent, increase the bid.}
\Else
\State $\nabla \mathcal L_t(b_t) \gets - C_t \cdot \frac{N_t}{1\,000 \ S_t} \left\{\frac{\beta}{b_t + \beta} \ \theta\!\left(\tau - \frac{S_t}{B_t}\right) - \frac{\beta}{b_t} \left[1 - \frac{\beta}{b_t} \ln \left(1 + \frac{b_t}{\beta}\right)\right]\right\}$
\EndIf
\EndProcedure

\Procedure{\textsc{apply Nadam to update the bids}}{see Figure~\ref{alg:nadam}}
\EndProcedure

\Procedure{\textsc{bound the bids in the correct interval}}{}
\State $b_{t+1} \gets \max\left( l_b, b_{t+1}\right)$
\State $b_{t+1} \gets \min\left( b_{t+1}, u_b\right)$
\EndProcedure

\State {\bf{return}} $b_{t+1}$
\EndFunction
\end{algorithmic}
\hrule
\caption{Algorithm for setting the base bid}
\label{alg:bid}
\end{figure}

\begin{figure}[!h]
\centering
\vspace{1pt}
Nadam algorithm
\hrule
\begin{algorithmic}[1]
\Function{\textsc{Nadam}}{}
\Require $\alpha_t; \mu_t, \mu_{t+1}; \nu; \epsilon;$
\Comment Hyperparameters
\Require $\vec m_t; \vec n_t;$
\Comment first/second moment vectors from the previous iteration
\Require $\vec b_t; \vec g_t;$
\Comment bids and gradient of the loss function
\State $\vec m_{t+1} \gets \mu_t \vec m_t + (1 - \mu_t) \vec g_t$
\State $\vec n_{t+1} \gets \nu_t \vec n_t + (1 - \nu_t) \vec g_t^2$
\State $\vec{\hat m} \gets \dfrac{\mu_{t+1} \vec m_t}{1-\prod_{i=0}^{t+1} \mu_i} + \dfrac{(1 - \mu_t) \vec g_t}{1-\prod_{i=0}^t \mu_i}$
\State $\vec n_{t+1} \gets \nu_t \vec n_t/(1-\nu^t)$
\State $\vec b_{t+1} \gets b_t - \dfrac{\alpha_t}{\sqrt{\vec n_t + \epsilon}} \vec m_t$
\State {\bf{return}} $\vec b_{t+1}$
\EndFunction
\end{algorithmic}
\hrule
\caption{One iteration of Nadam, adapted from \cite{Dozat2016}}
\label{alg:nadam}
\hrule
\end{figure}

\subsection{The analysis of the CPC}
The CPC can be rewritten as:
\begin{equation}
\text{CPC}_t = \frac{\text{CPM}_t}{1\,000 \ \text{CTR}_t},
\end{equation}
where CPM is the average Cost Per Mille, that is, the average price to pay for a thousand impressions, 
and CTR is the Click-Through Rate that we have already introduced.

First of all, we assume the CTR to be independent of the base bid and we estimate it from the market data as:
\begin{equation}
\text{CTR}_t = \frac{C_t}{N_t},
\end{equation}
where $N_t$ is the number of impressions bought and $C_t$ are the clicks.
The assumption of independence is justified by the fact that, if the media object filters are accurately set, all elements that are accessible by a single media object should be equally valuable.
Also, a correlation between CTR and bid would mean that there is a general consensus on what is the most promising impression to buy no matter the campaign advertisers are running.
The truth, as usual, lies in the middle: There is a certain correlation between CTR and bid, but in absence of better methods to estimate it we neglect it.
Introducing it in later improvements of the method will only require adding a term to the bid loss function.

We now have to find the CPM.
Following \cite{Zhang2014}, let us assume that the probability of winning an auction with a base bid of $b$ is given by an expression of the form:
\begin{equation}
P(b) = \frac{b}{b + \beta}, 
\label{eq:winning_prob}
\end{equation}
where $\beta$ is the median winning bid over all the inventory (since bidding $\beta$ gives a 50\% probability of winning).

We can define a probability density function:
\begin{equation}
p(b) = \frac{dP(b)}{db} = \frac{\beta}{\left(b + \beta\right)^2},
\end{equation}
that gives the percentage of inventory whose winning bid is exactly $b$.

RTB auctions often employ a second-price model to enforce truthful bidding \cite{vickrey1961counterspeculation,myerson1981optimal, edelman2007internet}. 
In such a situation, and remembering that the bid is typically expressed in total offer per thousand impressions, the CPM is given by:
\begin{equation}
\text{CPM}_t = \frac{\int_0^{b_t} x \, p(x) \, dx}{\int_0^{b_t} p(x) \, dx} = \beta_t \left[\left(1 + \frac{\beta_t}{b_t}\right) \ln \left(1 + \frac{b_t}{\beta_t}\right) - 1\right].
\label{eq:loss_cpm}
\end{equation}
We can notice the logarithmic increase of the average CPM at infinite bids representing competitors placing extremely high bids to acquire inventory, a strategy that gets rarer and rarer with increasing bids.
From Equation~\ref{eq:loss_cpm} we can estimate the value of the parameter $\beta_t$ by comparing the estimated value of the CPM with the actual CPM returned from the market during that epoch.

The CPC as a function of all the basic quantities of the problem then reads:
\begin{equation}
 \text{CPC}_t = \frac{N_t}{1\,000 \ C_t} \cdot \beta_t \left[\left(1 + \frac{\beta_t}{b_t}\right) \ln \left(1 + \frac{b_t}{\beta_t}\right) - 1\right].
\end{equation}
To obtain the derivative of the CPC part of the loss function we thus need only derive the CPM in (\ref{eq:loss_cpm}).
Calculations lead to:
\begin{gather}
\frac{d\text{CPM}_t}{db_t} = \frac{\beta_t}{b_t} \left[1 - \frac{\beta_t}{b_t} \ln \left(1 + \frac{b_t}{\beta_t}\right)\right];\\
\frac{d\text{CPC}_t}{db_t} = \frac{N_t}{1\,000 \ C_t} \frac{\beta_t}{b_t} \left[1 - \frac{\beta_t}{b_t} \ln \left(1 + \frac{b_t}{\beta_t}\right)\right].
\end{gather}

\subsection{The analysis of the amount of money spent}
In Equation~\ref{eq:winning_prob} we have made an assumption about the probability of winning an auction based on the base bid $b_t$ that is well evidenced.
We can try to leverage this assumption to find a relationship between $b_t$ and $S_t$.
Let us divide our discussion in two parts: The case of under-delivery and the case of correct delivery.

In the case of under-delivery, a media object buys the entire inventory that is available to it (because if more was available, it would buy it with the remaining money).
This quantity can be estimated with 
\begin{equation}
N_t(b_t) = I_{\text{tot},t} \cdot P(b_t),
\label{eq:total_inventory}
\end{equation}
where $I_\text{tot}$ is the total amount of inventory that would be available with an infinite bid.
The total money spent is then given exactly by:
\begin{align}
\label{eq:loss_spent}
S_t &= \frac{I_{\text{tot}, t}}{1\,000} \ \int_0^{b_t} p(x)\, x\, dx \\
&= \frac{I_{\text{tot}, t}}{1\,000} \ \beta_t \left[\ln\left(1+\frac{b_t }{\beta_t}\right) - \frac{b_t}{b_t+\beta_t}\right],\nonumber
\end{align}
where the factor 1\,000 comes from the fact that the bid are expressed in offer per thousand impressions.

The derivative of (\ref{eq:loss_spent}) with respect to the bid is given by:
\begin{equation}
\frac{dS_t}{db_t} = \frac{I_{\text{tot}, t}}{1\,000} \ \frac{b_t \cdot \beta_t}{\left(b_t + \beta_t\right)^2} = \frac{N_t}{1\,000} \frac{\beta_t}{b_t + \beta_t}.
\end{equation}
We could have found the first equality also applying the fundamental theorem of calculus to (\ref{eq:loss_spent}), while the second equality comes from the substitution $I_{\text{tot},t} = N_t(b_t) /P(b_t)$ (see Equation~\ref{eq:total_inventory})
which does not depend on the base bid $b_t$ because the dependences of $N_t$ and $P$ cancel each other.

In case of good delivery, instead, some pieces of inventory are not bought by the media object.
A change in the base bid would most probably modify the number of such pieces of inventory but won't change the total amount of money spent.
Therefore, in this case, $S_t$ is constant with respect to $b_t$ and its derivative is 0.

In order to discriminate between the two delivery regimes, we use a Heaviside step function $\theta\!\left(\tau - \frac{S_t}{B_t}\right)$,
where $\tau$ is an under-delivery threshold, typically set to 0.95 and not to 1 because a small amount of under-delivery is inherent to the discreteness of the problem.

\subsection{Proposed loss function and gradient}
We can now give the gradient with respect to the bids of the loss function $\mathcal L_t$ proposed in (\ref{eq:loss_bid}).
It reads:
\begin{equation}
\nabla \mathcal L_t(b_t) = - C_t \left(\frac{1}{S_t}\frac{dS_t}{db_t} - \frac{1}{\text{CPC}_t}\frac{d\text{CPC}_t}{db_t}\right),
\end{equation}
which, with the results found so far, becomes
\begin{align}
&\nabla \mathcal L_t(b_t) = - C_t \cdot \frac{N_t}{1\,000 \ S_t} \times\\
&\times\left\{\frac{\beta_t}{b_t + \beta_t} \ \theta\!\left(\tau - \frac{S_t}{B_t}\right) - \frac{\beta_t}{b_t} \left[1 - \frac{\beta_t}{b_t} \ln \left(1 + \frac{b_t}{\beta_t}\right)\right]\right\}.\nonumber
\end{align}

We notice that this equation is always well-defined, except when $S_t=0$.
This can happen in two situations: either there is no budget assigned to the strategy, in which case we impose no changes to be made since they would have no effect anyway;
or there is a budget assigned but the strategy doesn't manage to spend anything, in which case we are probably seriously underbidding and we fix the value of the gradient to be negative.

With this loss function, we perform a Nadam gradient descent \cite{Dozat2016, Kingma2014} and then bound the result to be in between a minimal and a maximal bid set by the client.
Unlike in budget partitioning, we choose an additive gradient descent because we don't need any normalization.

As a last remark on the base bid setting, there is currently a resurgence in first price auctions.
Our method is still applicable even in this situation, provided a few changes are made to the form of the equations.
In particular, (\ref{eq:loss_cpm}) and (\ref{eq:loss_spent}) would read respectively:
\begin{gather}
\text{CPM}_t = b_t;\\
S_t = \frac{I_{\text{tot}, t}}{1\,000}\  b_t \int_0^{b_t} p(x) \, dx,
\end{gather}
giving rise to different, but nevertheless well-defined update rules.

Recently, another research paper that deals with the bidding algorithm was published \cite{Ren2018}.
While there are similarities between their approach and ours, we chose to maximize directly the number of clicks instead of defining another utility function that needs other hyperparameters such as the monetary value of each click.
Moreover, the method we propose in this paper does not need to have one data point per impression (an information that we assume is not at our disposal) but only the average over a certain period of time.

%% file: pacing.tex
\section{\skott: Pacing control}
\label{sec:pacing}

The third and last sub-routine is the one that controls the delivery ratio.
It checks that the total amount of money spent in the campaign so far follows the desired profile.
If that's not the case, it increases the total budget available for the next epoch.
Notice that our goal is not to determine what is the best delivery profile of the campaign over time, but only to stick to it as well as possible.
This sub-routine is the simplest one since it only sets a single scalar parameter, unlike the previous two who sets a vectorial one.

\begin{figure}[!ht]
\hrule
\centering
\vspace{1pt}
Algorithm for pacing control
\hrule
\begin{algorithmic}[1]
\Function{\textsc{pacing control}}{} 
\LeftComment{$t$: the current epoch}
\LeftComment{$T-t$: epochs left}
\LeftComment{$\mathcal S_t$: actual spent until $t$}
\LeftComment{$\bar{\mathcal S_t}$: ideal spent until $t$}
\LeftComment{$\bar{\mathcal B}_{t+1}$: ideal budget for the next epoch} 
\LeftComment{$\eta$: aggressiveness parameter}

\State $\Delta \mathcal S_t \gets \bar{\mathcal S_t} - \mathcal S_t$
\State $\Delta \mathcal B_{t+1} \gets \eta \cdot \Delta \mathcal S_t / (T-t)$
\State $\mathcal B_{t+1} \gets \bar{\mathcal B}_{t+1} + \Delta \mathcal B_{t+1}$
\State {\bf{return}} $\mathcal B_{t+1}$
\EndFunction
\end{algorithmic}
\hrule
\caption{Algorithm for pacing control}
\hrule
\label{alg:pacing}
\end{figure}

Typically, advertisers want to control exactly how much money they spend during the campaign.
For example, the simplest delivery profile is the uniform one, where the ideal amount of money spent until $t$ is equal to the total budget of the campaign times the fraction of the campaign that has elapsed already.
However, the money that was really spent on the market doesn't always correspond to the desired amount: unforeseen technical issues, fluctuations in the available inventory, sudden changes in the properties of the market; they all can contribute to a variation in the amount of money spent, typically resulting in under-delivery.

Before the budget partitioning and base bid setting sub-routines can react to the under-delivery and adapt their parameters, the actual delivery of the campaign will have lost ground to the ideal one.
It is desirable then to take some measures in order to catch up with the ideal spent as soon as possible.

The hourly budget $\mathcal B_{t+1}$ set by the algorithm looks like this:
\begin{align}
\mathcal B_{t+1} &= \bar{\mathcal B}_{t+1} + \Delta \mathcal B_{t+1},\\
\Delta \mathcal B_{t+1} &= \frac{\eta}{T-t} \left(\bar{\mathcal S_t} - \mathcal S_t\right),
\end{align}
where $\bar{\mathcal B}_{t+1}$ is the ideal hourly budget, 
$\bar{\mathcal S}_t$ and $\mathcal S_t$ are respectively the ideal and actual amount of money spent until epoch $t$, 
$T-t$ is the number of epochs left,
and $1 \leq \eta \leq T-t$ is the aggressiveness parameter.
If the aggressiveness is set to 1, the algorithm tries to evenly spread the correction over the rest of the campaign.
Surprisingly, this is not good: the reason is that a small amount of under-delivery is very common and it won't be contrasted fast enough, imposing a money rush toward the end of the campaign.
Also, we typically want to regain the ideal spend curve at a higher speed.
However, too large a value of aggressiveness is not desirable either, because it could mean a very large sudden injection of money, possibly reducing the quality of our inventory and breaking the simple assumptions we had to make to construct a model.
We typically choose values between 2 and 20.

A schematic view of the algorithm is presented in Figure~\ref{alg:pacing}.

%% file: experiments.tex
\section{Experimental results}
\label{sec:experimental_results}

We tested our algorithm on a simulated environment.
(More on the characteristics of the market simulator in \ref{app:market}.) 
We will show the results as follows:
First we will compare different budget partitioning algorithms while keeping no optimization on the base bids or on the pacing.
Then, we will compare base bid setting algorithms on top of the budget partitioning we presented in this paper.
Finally, we will show the advantages of introducing the pacing control algorithm on top of the budget partitioning and base bid setting algorithm we chose.

\subsection{The comparison of budget partitioning algorithms}
We compared our algorithm to three other algorithms: 
(1) A \emph{vanilla} algorithm (codenamed \texttt{vnl}) that does absolutely nothing.
(2) A multi-armed bandit algorithm inspired by \cite{Auer2002,Besbes2014}, codenamed \texttt{mab}.
(3) A linear optimization algorithm inspired by \cite{Chen2011}, codenamed \texttt{lop}, that maximizes the clicks under the constraints of the total available budget and an interval of admitted budgets for every media object.
More information about these algorithms can be found in \ref{app:algorithms}.

\begin{figure}
\centering
\includegraphics[width = \linewidth]{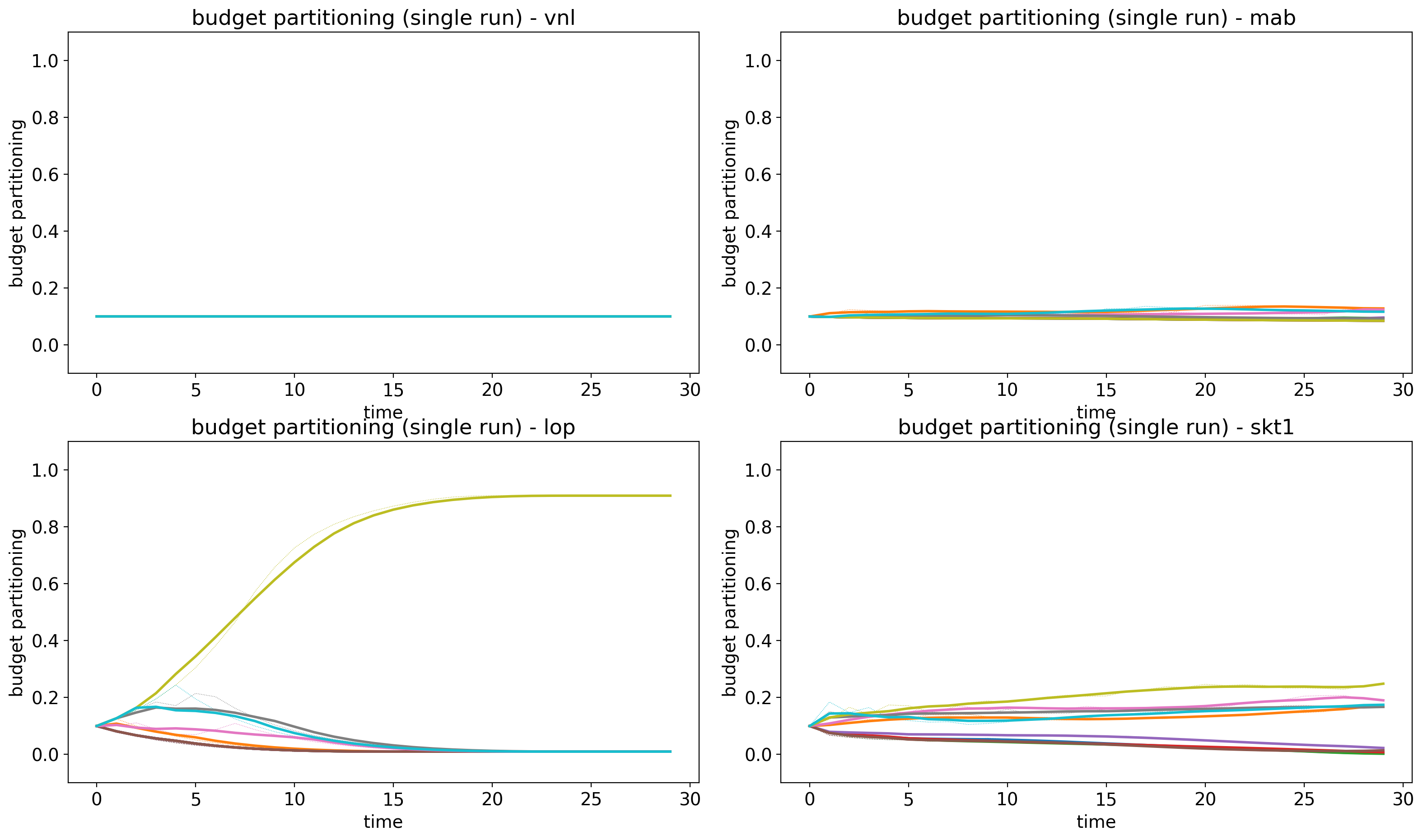}
\caption{The evolution of the budget partitioning for 10 media objects, according to the four different algorithms we test, taken over a single run of the optimization.
The solid lines show a smoothed version of the dotted lines for better visualization.}
\label{fig:br_single}
\end{figure}

\begin{figure}
\centering
\includegraphics[width = \linewidth]{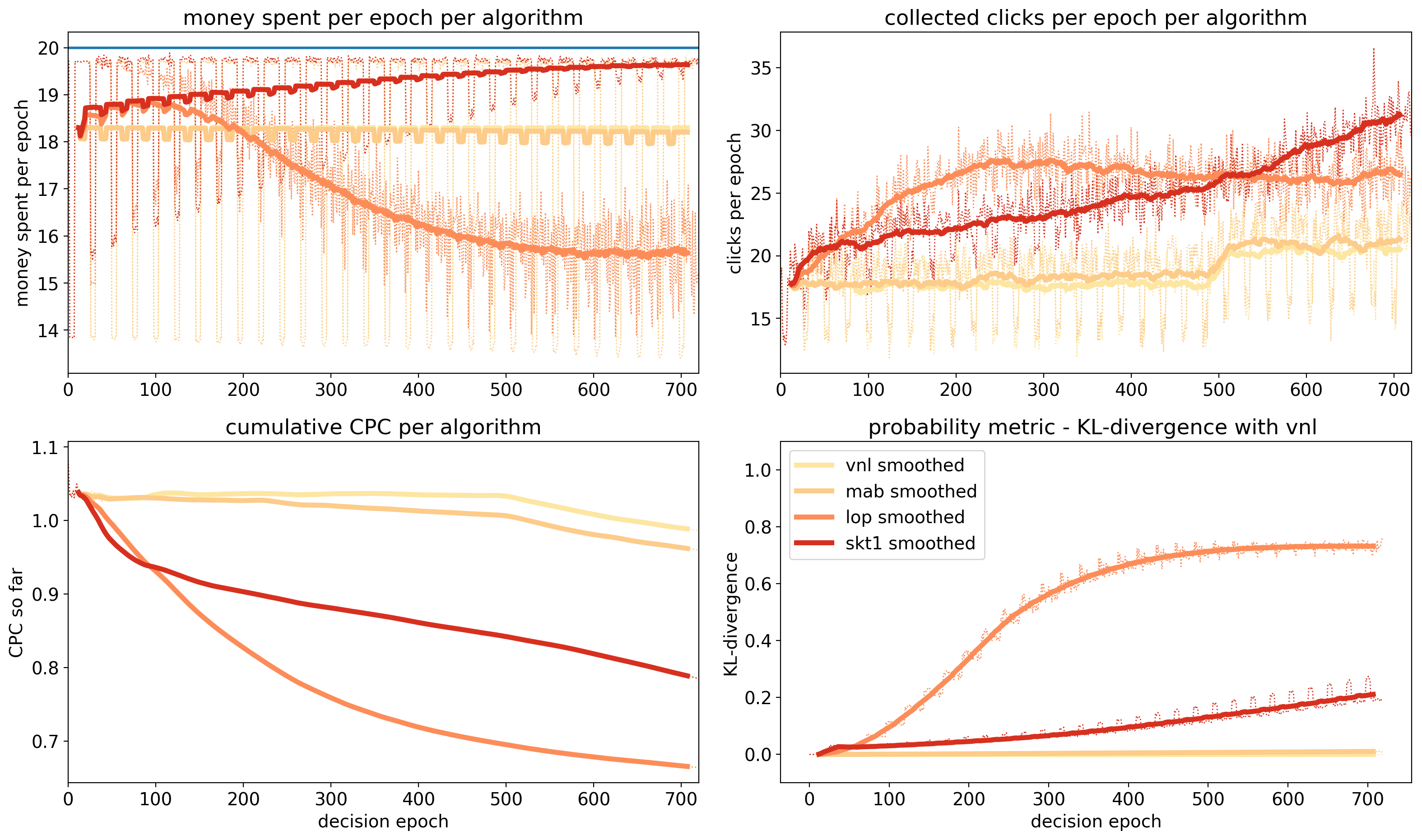}
\caption{The four metrics we use to evaluate the algorithms: money spent, clicks obtained, CPC, and divergence from the uniform distribution as obtained from an average of 20 optimizations on different randomly chosen starting points. 
The solid lines show a smoothed version of the dotted lines for better visualization.}
\label{fig:epoch_br}
\end{figure}

\begin{table}
\centering
\caption{Optimization Results for Budget Partitioning}
\begin{tabular}{c|c|c|c|c}
\textbf{algo} & \textbf{spt} & \textbf{clk} & \textbf{cpc} & \textbf{kld}\\
\hline
\texttt{vnl}  & 91.2 \% & 100.0 \% & 0.990 & 0.000\\
\texttt{mab}  & 91.0 \% & 102.6 \% & 0.979 & 0.005\\
\texttt{lop}  & 84.5 \% & 137.7 \% & 0.664 & 0.503\\
\texttt{skt1} & 96.2 \% & 132.5 \% & 0.785 & 0.095\\

\end{tabular}
\label{tab:budget_results}
\end{table}

Figure~\ref{fig:br_single} gives the comparisons of these algorithms.
We present a simulation with day parting, i.e., with a different algorithm running for each hour of the day.
The total number of epochs is 30, the number of days in a month.
The first thing we want to point out is that \texttt{lop} is quite greedy, as we expected, while \texttt{mab} is almost like \texttt{vnl}. 
This is due to the fact that only one media object per epoch is updated, giving just 30 small kicks to the initial situation.
Our proposed algorithm, \texttt{skt1}, seems to strike in between:
It moves quickly without becoming greedy.

To quantify this result, we measure greediness by calculating the KL-divergence \cite{Kullback1951} of the proposed budget repartition with respect to the uniform distribution $\vec u$.
The KL-divergence is a widely used method: for example, in reinforcement learning, it measures the distance between two policies, i.e., two different courses of action that optimize a given reward \cite{Auer2010,Schulman,Plappert2017}.
It is defined as:
\begin{equation}
\Delta(\vec w, \vec u) = \sum_i w_i \log\left(\frac{w_i}{u_i}\right).
\label{eq:kl}
\end{equation}
A value of $\Delta(\vec w, \vec u) = 0$ means that the distribution of the weights is exactly the same as $\vec u$.
On the other hand, the maximal value is obtained by a greedy distribution $\vec w_g^{(i)}$ where one of the elements is 1 and all the others are 0. In this case, the KL-divergence measures $\log(K)$.
The values in the lower right plot of Figure~\ref{fig:epoch_br} are values of KL-divergence rescaled by a factor $\log(K)$, so that they are always constrained between 0 and 1 independently of the number of media objects.

These qualitative discussions find their quantitative conclusions in Figure~\ref{fig:epoch_br} and in Table~\ref{tab:budget_results}:
The first column of numerical values (\textbf{spt}) represents the percentage of the initial budget that was spent, the second (\textbf{clk}) the additional clicks in percent with respect to the \texttt{vnl} algorithm, the third (\textbf{cpc}) the total CPC of the campaign, and the fourth (\textbf{kld}) the distance of the budget repartition from the uniform distribution.

If one considers only the total number of assigned clicks as the metric to measure the performance of the algorithms, \texttt{lop} wins over \texttt{skt1} by a small margin.
Also, its total calculated CPC is slightly lower, meaning that every click costs less money.
However, on the bottom right panel of Figure~\ref{fig:epoch_br}, we can see how greedy \texttt{lop} is.
This reflects on the total amount of money spent on the top left panel: if the desired media object doesn't have enough available inventory, this algorithm can not react decisively.
Nothing grants us that this situation won't happen in real life with even more damaging results, in particular, a severe under-delivery of the budget.
On the contrary, \texttt{skt1} manages to spend almost all the available budget (represented by the purple line) even without an explicit optimization on the bid and the pacing.
The increased adaptability makes it more resistant to the real market test and thus more valuable.

\subsection{The rest of the analysis}
We now choose the \texttt{skt1} algorithm for the budget partitioning and study the effect of different base bid setters.
This time, we compare the new \texttt{skt2} algorithm to other two algorithms: again \texttt{vnl} where bids are not changed, and also \texttt{pst}, an algorithm that uses a predetermined set of rules. 
We also add in the analysis the comparison of \texttt{skt2} on top of \texttt{skt1} with the full \skott\ algorithm, which also contains the pacing control sub-routine.

Let us explain first a little bit about \texttt{pst}.
The predetermined set of rules analyzes the CPC first: if it is higher than a certain goal set at the beginning of the campaign, the bid is reduced by a certain fixed multiplicative constant unless the media object is under-delivering. In that case \texttt{pst} will still try to slightly increase the bid.
We can see many inconveniences in this approach compared to the algorithm we presented in Section~\ref{sec:bid}: first of all we need an additional external parameter, that is, the goal CPC.
Furthermore, it makes very little use of the data from the market, reducing the adaptability.

\begin{figure}
\centering
\includegraphics[width = \linewidth]{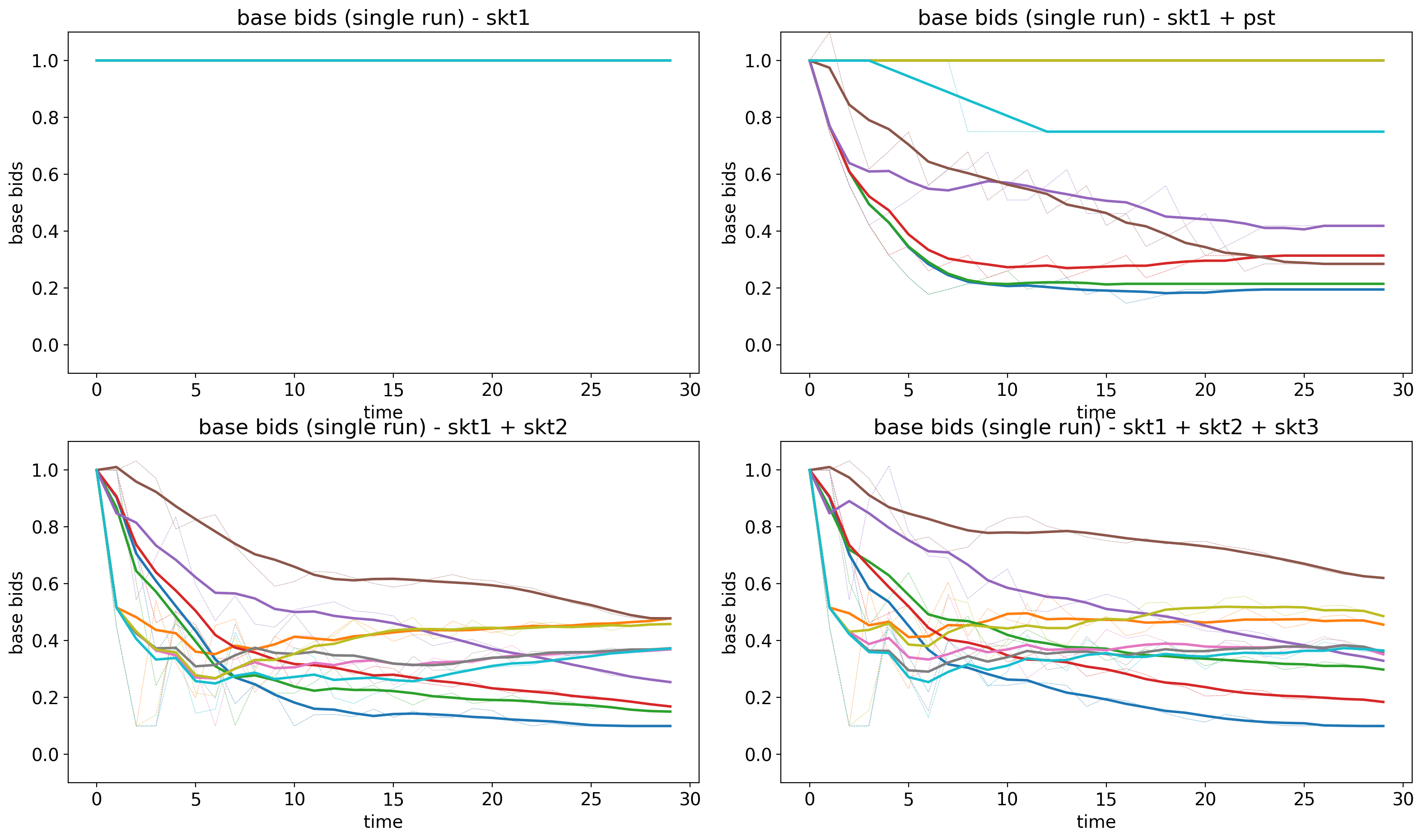}
\caption{The evolution of the base bid for 10 media objects, according to the four different algorithms we test, taken over a single run of the optimization.
The solid lines show a smoothed version of the dotted lines for better visualization.}
\label{fig:bid_single}
\end{figure}

\begin{figure}
\centering
\includegraphics[width = \linewidth]{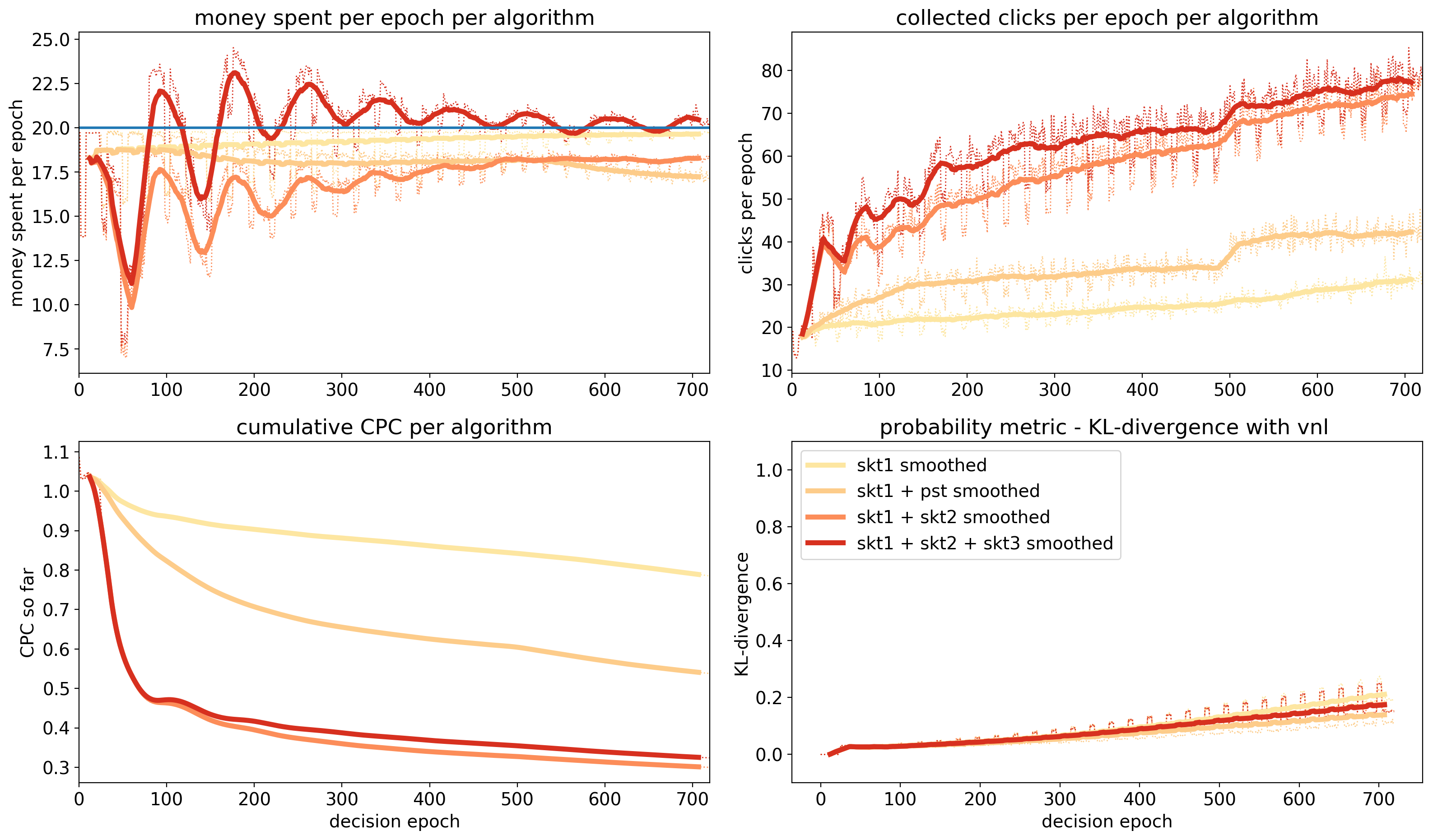}
\caption{Same as Figure~\ref{fig:epoch_br}, but for the base bid setting algorithms. The solid lines show a smoothed version of the dotted lines for better visualization.}
\label{fig:epoch_bid}
\end{figure}

\begin{table}
\centering
\caption{Optimization Results for Base Bid and Pacing.}
\begin{tabular}{c|c|c|c|c}
\textbf{algo} & \textbf{spt} & \textbf{clk} & \textbf{cpc} & \textbf{kld}\\
\hline
\texttt{skt1}               & 96.2 \% & 100.0 \% & 0.785 & 0.095\\
\texttt{skt1 + pst}         & 90.2 \% & 136.9 \% & 0.538 & 0.071\\
\texttt{skt1 + skt2}        & 84.8 \% & 231.0 \% & 0.300 & 0.086\\
\texttt{skt1 + skt2 + skt3} & 99.8 \% & 251.6 \% & 0.324 & 0.085\\
\end{tabular}
\label{tab:bid_results}
\end{table}

These limitations have an effect on the results, as can be seen from Figure~\ref{fig:epoch_bid} and from Table~\ref{tab:bid_results}.
Differently from before, the baseline for the column (\textbf{clk}) is now \texttt{skt1} and not \texttt{vnl}.
The \texttt{skt2} algorithm manages to outperform both \texttt{vnl} and \texttt{pst} by a vast amount, obtaining a larger number of clicks while spending less money.

From the top left corner of Figure~\ref{fig:epoch_bid} one can see that the amount of money spent oscillates quite a bit at the beginning of our proposed algorithm.
We see this stems from a similar oscillation in the algorithm's attempts to find the appropriate bids, as can be seen from Figure~\ref{fig:bid_single}: the peaks of money spent correspond to bids slightly higher than the optimal and vice versa.
Finally, we see that adding the third and last part of the algorithm manages to spend almost the entire initial budget, obtaining a small increase in clicks at the expense of a slightly higher CPC.

It is interesting to notice that the last plot, containing the KL-divergence, shows that the budget repartition changes slightly depending on which bidding algorithm we choose.
This is understandable because, by changing the base bid, we actually modify the perceived quality of the media object and thus generate different inputs for the different iterations of the budget partitioning algorithm.
However, these modifications are small enough to be neglected at first order.

%% file: conclusion.tex
\section{Conclusion}
\label{sec:conclusion}

We have introduced a method for advertisers to optimize the management of Demand Side Platforms when running an advertising campaign composed of many separate media objects.
The method, that we call \skott\ algorithm, is an iterative method that makes only a few general assumptions on the mathematical model of the market.
We present it here applied to a campaign for the optimization of the number of generated clicks.

The \skott\ algorithm is composed of three complementary parts: 
Firstly, the best partitioning for the budget across all media objects is calculated. 
This is achieved by estimating the quality of each media object and trying to obtain the maximum number of clicks through an exponentiated gradient descent method.
Second, the best base bid for each media object is calculated.
Here we use the assumption and corresponding evidence from \cite{Zhang2014} that relates the bid to the probability of winning the corresponding auction. We expand on this assumption to propose a model relating variation in bids to variation in the number of clicks obtained by each media object.
We finally apply a Nadam technique \cite{Dozat2016, Kingma2014} on the market data to find the best base bids for maximizing the number of clicks.
The third and last part determines the amount of budget to use at every epoch in order to stay as close as possible to the desired spend profile.

The proposed algorithm has been tested on a simulated environment that we created for the occasion and that we present in the appendices.
Under these circumstances, the proposed algorithm gives impressive results, more than doubling the total amount of obtained clicks in the considered experiments.

%% file: market.tex
\section{Model of the market}
\label{app:market}

We created a back-test platform for analyzing different campaign management algorithms.
The work-flow of the platform is conceptually divided in five steps:
\begin{enumerate}
\item Parameters are chosen that describe the problem we have at hand.
\item Data is created using these parameters.
\item Loss functions are chosen.
\item All algorithms are launched independently. They work on the same data and their goal is to maximize the total number obtained during the campaign.
\item The results of the different algorithms are compared: we plot budget repartition, base bids, greediness of the algorithms, cumulative CPC, spend profiles, and collected clicks over time.
\end{enumerate}
In this section, we will discuss the first two steps, that relate to the creation of the back-test platform itself.
Step 3 and 4 are discussed in Section~\ref{sec:budget}, \ref{sec:bid}, and \ref{sec:pacing} of the main body of the article, while some plots of the results are presented in section~\ref{sec:experimental_results}.

\subsection{Choice of the parameters}
There are just a few important parameters for the problem. They can be chosen independently, but the quality of the output policy heavily depends on their interaction.
These parameters are:
\begin{itemize}
\item The total number of epochs $T$. This is fixed by the length of the campaign and by the duration the data batches. Since we consider hourly data and a campaign lasts for one month, we typically work with $T = 24*30 = 720$.
\item The day parting boolean: If true, we only consider data that comes from the same hour of the day, effectively launching 24 parallel algorithms and reducing the number of epochs per algorithm by a factor of 24. The reasons behind this are further explored in \ref{app:periodicity}.
\item The number of media objects $K$. Typically, a larger $K$ increases the probability of having good media objects but also increases the cost of exploration. The optimal number of media objects, therefore, depends on the budget at our disposal.
\item The $\text{CPC}_\text{goal}$, a business-driven parameter that says how much money the advertiser wants to spend on average to get a click. We use this only for the \texttt{pst} algorithm.
\item The total campaign budget $\mathcal B$ or, equivalently, the average budget per epoch per media object $\text{bem} = \mathcal B / (T \cdot K)$. In particular, there is a sort of phase transition at the critical value $\text{bem} = \text{CPC}_\text{goal}$: Above that value, all media objects should statistically be able to obtain at least one click per epoch, making for efficient exploration even in the first epoch alone, when no optimization has started yet. Below that, clicks become rare and many epochs are needed to efficiently understand the relative quality of the media objects.
\item The number of repetitions of the experiment, $E$. The process of determining the best algorithm heavily depends on the input data, which is randomly generated. A single sample from a random generation might have some characteristics that are more convenient for a particular algorithm. Taking several samples of data and averaging the results of the numerical calculations over all the samples reduces the impact of this error at the cost of a linear increase in computational time. Typically we take $E=20$.
\end{itemize}

\subsection{Data creation}
Data creation is handled by a Python class. Given the budgets $\vec B_t$ and the base bids $\vec b_t$ of all media objects for epoch $t$, it returns the impressions bought $\vec N_t$, the clicks obtained $\vec C_t$, and the money spent $\vec S_t$ in that epoch for all media objects.
To make synthetic data closer to real data, the class is constructed to mimic the structure of the market data we can access.
Therefore, it takes as an input a subset of the same parameters that we send to the market during real-life campaigns. 
Also the output is made of a subset of the same results we obtain from the market in real life.

The underlying model of the market is exactly the same as described in section~\ref{sec:bid}.
For the creation of data, we generate the CTRs, the total inventory available on the market $\vec I_\text{tot}$, and the median winning bids $\vec \beta$ separately for each media object, sampling from appropriate intervals uniformly at random.
Then we use (\ref{eq:loss_cpm}) to calculate the average CPM of the media objects as a function of their bids, $\vec b_t$.
With all this, we can find out how many impressions each media object will buy at epoch $t$ by using:
\begin{equation}
\vec N_t = \min\left(\frac{1\,000 \ \vec B_t}{\vec{CPM}_t}, \ \vec I_{\text{tot}, t} \cdot \vec P\!\left(\vec b_t\right) \right),
\end{equation}
where the first element is the number of impressions one would buy with infinite available inventory,
and the second one is the total available pieces of inventory with bid $\vec b_t$.

We finally use the equation:
\begin{equation}
\vec S_t = \vec N_t \cdot \frac{\vec{CPM}_t}{1\,000}
\end{equation}
to find out how much money each the media object spent and, consequently, how much it under-delivered.

Once the purchase of impressions is done, the market simulator assigns the clicks by sampling from a binomial distribution with probability of positive outcome given by the CTR of each media object.

The market simulator allows for a certain amount of control on the variation over time of the quality of the media objects, since all the quantities we generate can be modified at any epoch. 
This is a very important feature of the data: a model with static media objects tends to select algorithms that don't explore, while dynamic media objects require exploration to keep track of the changing environment.
Different time dependencies, therefore, lead to different best algorithms.
More details about sources of periodicity can be found in \ref{app:periodicity}.

%% file: periodicity.tex
\section{Dealing with time variations}
\label{app:periodicity}

As already mentioned in \ref{app:market}, there can be variations in the quality of media objects with time.
While the causes are various, the main effect is the double periodicity induced by the day-night cycle and the weekly cycle \cite{Yuan2013}. 
In particular, there is a big drop in the volume of impressions dealt and the number of clicks at night that often leads advertisers to forgo buying impressions at these times.

But variations are not always periodic, nor globally affecting all the media objects at the same time.
A change in the relative quality of the media objects can happen over time due to external factors.
A typical example could be a media object advertising a live event: the distance in time from the day of the event is an important parameter for users to decide whether to buy a ticket.

Finally, some changes might be due to correlations not considered in our model. 
For example, a change in the base bid that we offer during auctions might lead to a modification of the CTR of the impressions we are able to buy. 

The solution we take in order to deal with such issues is to launch 24 different algorithms, one for every hour.
The advantage is clear: in case some media objects are turned off at a certain moment of the day, they won't affect the perceived quality of the media object during other hours.
However, there are obvious disadvantages as well: we discard data that might still give valuable information and convergence is 24 times slower.

For the aperiodic modifications over time, we want to have fast responses to the changes in the market.
Since information on the changes is obtained through the purchase of impressions, we try to keep the algorithm as far from greedy as possible while still increasing the number of obtained clicks.

%% file: dataProcessing.tex
\section{Data pre-processing}
\label{sec:data_pre-processing}
In the real world, when we receive the data, there are often missing values that need to be filled before starting the optimization.
Here, we fill the missing values using a combination of three different approaches: 
a) backward filling; where  we propagate the next valid observation,
b) linear interpolation method; which is a method of curve fitting using linear polynomials to construct the missing values,
and c) weighted moving averaging approach; which is an averaging that has multiplying factors to give different weights to data points at different positions.

Let \textbf{x} = [$x_1$, $x_2$, {\sc nan}, $x_4$, ..., $x_t$, {\sc nan}, ..., {\sc nan}] be of assumed length $\tau$. 
As one can see, there are some {\sc nan}s before epoch $t$, and a series of {\sc nan}s between $t$ and $\tau$.
In the case of having some missing values at the beginning of the vector, we fill the {\sc nan}s using the backward filling method. 
For instance, [{\sc nan}, $x_i$, $x_j$] gives [$x_i$, $x_i$, $x_j$].

Next, we fill the {\sc nan}s up to epoch $t$ using a linear interpolate method.
As a result, there will be no missing values between the epochs 1 and $t$.

Last, we fill the {\sc nan}s between the epoch $t$ and $\tau$ using a weighted moving averaging approach.
Let us consider \textbf{x'} = [$x_1$, ..., $x_t$, {\sc nan}, ..., {\sc nan}], where for all $1 \leq i \leq t$,  $x_i \neq$ {\sc nan}, and for all $t+1 \leq i \leq \tau$, $x_i = $ {\sc nan}. 
To do so, we consider a weight vector in the filling process.
In an $t$-index weighted moving average, the latest data point has weight $t$, the second latest $t- 1,\dots$, terminating at one.
Therefore, the estimation of data point $t+1$ is defined as:
\begin{equation}
x_{t+1} = \displaystyle \frac{( 1 \cdot x_1 + 2 \cdot x_2 + ... + (t-1)\cdot x_{t-1} + t\cdot x_t )}{( t \cdot (t+1)/2 )} 
\label{eq:wma}
\end{equation}

All the missing values $x_i$ for $t+1 \leq i \leq \tau$ are filled using (\ref{eq:wma}).

%% file: algorithms.tex
\section{A quick glance at the competing algorithms}
\label{app:algorithms}

In section~\ref{sec:experimental_results} we mention three algorithms that we use as a comparison against our own.
The \texttt{vnl} algorithm needs no explanation, because it corresponds to a non-optimized campaign in which the initial parameters are kept constant.
On the other hand, \texttt{mab} and \texttt{lop} deserve a few words, which we will spend in the next paragraphs.

\subsection{The multi-armed bandit algorithm}
The Multi-Armed Bandit problem (MAB) deals with an agent (that is, whoever or whatever is able to take an action, such as a person or a robot) which, at definite time moments $t$ known as epochs, is faced with a number $K$ of possible different actions, each leading to a different reward. 
The goal is typically to find the course of actions, also called a policy, that maximizes the received reward. 
The difficulty of the problem is that the rewards are not known in advance and can be stochastic; a certain number of trials is necessary in order to explore the system before understanding what is the optimal policy to exploit.

The exploration-exploitation dichotomy is at the heart of all multi-armed bandit algorithms: the information acquired during the trials comes at the cost of exploring unknown actions that might lead to poor immediate rewards.

Stochasticity in a policy assures the exploration: therefore, it is needed at the beginning and is often reduced over time to ensure eventual convergence to the best action. 
This is no longer valid when the rewards change over time: In that case, the exploration is always needed to keep track of the moving averages.

Mathematically speaking, the MAB is defined by a tuple $\langle A, R \rangle$, where $A$ is the set of all possible actions the agent can take and $R$ is the set of the rewards associated with those actions. 
The policy $\pi$ is a probability distribution over $A$: The next action is chosen by sampling over the policy $\pi$.

In our case, we want to choose the best media object to bid with on incoming inventory items.
The $K$ actions inside $A$ correspond to {``choose media object 1", ..., ``choose media object K"}.
However, the agent that chooses the action to perform is a bidding algorithm that is not directly under our control.
The only thing we can influence is the distribution of the results: 
If, for example, we want to enact a policy for which media object 1 should be preferred 80\% of times (exploitation) and all the others should be equally distributed among the remaining 20\% (exploration), we can set the budget associated to every policy accordingly. Since a media object can not place a bid unless it has enough available budget, this effectively forces the desired average behavior over a certain time period.

A second difference with the standard MAB problem relates to the information we have access to.
We don't have access to all the information given to the bidding algorithm, but only to aggregate information on winning impressions every hour, therefore, to the average reward of every media object at every epoch.
However, this is strictly related to our inability to access the bid itself and doesn't introduce any additional constraint.

In spite of these differences, we can still create an algorithm based on the multi-armed bandit problem.
To do so, we follow the algorithm \texttt{exp3} as presented in \cite{Auer2002}, copied in Figure~\ref{alg:exp3} for convenience.

\begin{figure}[!ht]
\hrule
\centering\vspace{1pt}
Budget partitioning with exp3
\hrule
\begin{algorithmic}[1]
\Function{\textsc{Exp3 budget partitioning}}{}
\LeftComment{$\gamma$: real number $\in (0, 1]$}
\Procedure{Initialization}{}
\State $w_i(1) = 1 $ for $i = 1, ..., K$
\EndProcedure
\Procedure{Budget repartitioning}{}
\For{$t = 1, 2...$}
\State  $p_i(t) \gets (1-\gamma) \displaystyle \frac{w_i(t)}{\sum_{j=1}^K w_j(t)} + \frac{\gamma}{K}  \qquad i = 1, ..., K$
\State draw $i_t$ randomly accordingly to the probabilities $p_1(t), ..., p_K(t)$.
\State receive reward $R_{i_t} (t) \in [0, 1)$
\For{$j = 1, ..., J$}
\begin{align}
\hat x_j(t) &\gets 
\begin{cases}
R_i(t) / p_j(t) & \text{if } j = i_t,\\
0 & \text{otherwise}
\end{cases}\\
w_j(t+1) &\gets w_j(t) \exp\left(\frac{\gamma \ \hat x_j(t)}{K}\right)
\end{align}
\EndFor
\EndFor
\EndProcedure
\EndFunction
\end{algorithmic}
\hrule
\caption{The \texttt{exp3} algorithm as presented in \cite{Auer2002}.} 
\hrule
\label{alg:exp3}
\end{figure}

The reward function we choose is:
\begin{equation}
\vec R = \frac{|\vec x|}{1+|\vec x|},
\qquad
\vec x = \frac{\vec C}{\text{C}_\text{goal}} * \frac{\text{CPC}_\text{goal}}{\vec{CPC}},
\end{equation}
where $\vec C$ is a vector of clicks and the goal (in terms of clicks) is given by the maximum of the running exponential average of the number of clicks obtained by the media object and the ratio between the allotted budget and the goal CPC.
This reward has been chosen because we value media objects that have low cost per goal and attract many clicks.
We don't use the variable $\vec x$ as a reward because we prefer to work with rewards bounded between 0 and 1. 

Notice that this algorithm, while certainly offering improvements over the \texttt{vnl} algorithm, is only using a fraction of all the information that we have since, conforming to the possibility of a real MAB problem, it uses the reward of a single media object at every epoch.

\subsection{The linear programming algorithm}
The problem of collecting the largest possible number of clicks can be written as an optimization under some linear constraints.
In particular, we try to maximize the number of clicks under the (external) constraint of the total budget and a (self-imposed) constraint on the amount of variation of the budget per each epoch.

\begin{figure}[h]
\hrule
\centering\vspace{1pt}
Linear programming algorithm for budget partitioning
\hrule
\begin{algorithmic}[1]
\Function{\textsc{Linear optimization budget partitioning}}{}
\LeftComment{$\vec{CPC}$: an estimation of the CPC of each media object}
\LeftComment{$\tilde{\mathcal B}_t$: the remaining budget in the campaign}
\LeftComment{$\vec{l}$: the lower bounds to the budget per media object}
\LeftComment{$\vec{u}$: the upper bounds to the budget per media object}

\Procedure{\textsc{budget partitioning}}{}
\State $R \gets \tilde{\mathcal B}_t$
	\Comment the remaining budget to allocate
\State $\vec B \gets \vec l$
	\Comment all media objects get at least the minimum allowed
\State $R \gets R - \sum_i B_i$
	\Comment start reducing the remaining budget
\State $\vec{idx} \gets $ indices that sort $\vec{CPC}$ in descending order
\For {i in $\vec{idx}$}
	\State $a_i \gets \text{min}(R, u_i - l_i)$ 
    	\Comment additional money to allocate to $B_i$
	\State $B_i \gets B_i + a_i$
    \State $R \gets R - a_i$
\EndFor
\EndProcedure

\State {\bf{return}} $\vec B$ 
\EndFunction
\end{algorithmic}
\hrule
\caption{Linear programming algorithm for budget partitioning} 
\hrule
\label{alg:lop}
\end{figure}

Let's assign to the variables $\vec B \in R^K$ the budgets of the $K$ media objects.
The function to maximize is:
\begin{equation}
f(x) = \vec{c}^\intercal \cdot \vec B,
\end{equation}
where $\vec{c} = 1/\vec{CPC}$ is a vector of coefficients that relates the budget to the expected number of clicks.
The vector $\vec{CPC}$ is estimated from real market data with the formulas:
\begin{gather}
\vec{\hat S}_t = \vec S_t + \gamma \vec{\hat S}_{t-1},
\qquad
\vec{\hat C}_t = \vec C_t + \gamma \vec{\hat C}_{t-1}, \\
\vec{CPC}_t = \frac{\vec{\hat S}_t}{\vec{\hat C}_t},
\end{gather}
where $\vec S_t$ and $\vec C_t$ are respectively the vector of the total money spent and the the total clicks gathered at a certain epoch $t$. The values 
$\vec{\hat S}$ and $\vec{\hat C}$ are initialized at zero when $t=0$ 
and $\gamma \in [0, 1]$ is the discount factor (see Section~\ref{sec:budget} for more details on the discount factor).

The constraints can be written in the form:
\begin{equation}
A * \vec B \leq \vec{v},
\qquad
A \in R(2k+1, k), 
\qquad
\vec v \in R(2k+1).
\end{equation}
The matrix $A$ has only ones in the first row followed by two $K \times K$ diagonal matrices containing respectively only $-1$ and only $1$.
The vector $\vec{v}$, instead, contains the remaining budget $\tilde{\mathcal B}_t$ as the first element followed by the $2\,K$ bounds to the budget to be assigned to each media object:
The first $K$ are the lower bounds $l_{i, t}$ (with a minus sign in front to reverse the direction of the inequality) 
while the second $K$ are the upper bounds $u_{i, t}$.
For example, in the case $K = 3$, they read:
\begin{equation}
A = 
\begin{pmatrix}
1 & 1 & 1\\
-1 & 0 & 0\\
0 & -1 & 0\\
0 & 0 & -1\\
1 & 0 & 0\\
0 & 1 & 0\\
0 & 0 & 1
\end{pmatrix},
\qquad
v = 
\begin{pmatrix}
\tilde{\mathcal B}_t\\
-l_{0, t}\\
-l_{1, t}\\
-l_{2, t}\\
u_{0, t}\\
u_{1, t}\\
u_{2, t}
\end{pmatrix}
\end{equation}
The values of the lower and upper bounds control are set as following:
\begin{equation}
l_{i,t+1} = \alpha_l \cdot S_{i, t},
\qquad
u_{i,t+1} = \alpha_u \cdot S_{i, t},
\end{equation}
where $\alpha_l < 1$ and $\alpha_u >1$.
These two quantities control how much variation is possible to introduce between epochs and are therefore related to the learning rate of the algorithm.

The solution to this system can be found analytically and is presented in Figure~\ref{alg:lop}. 
It is important to notice that, in case the estimation of the CPC is sufficiently constant, this algorithm converges exponentially to the greedy solution.
However, in our back-test simulations, stochasticity in the market and the use of the discount factor $\gamma > 0$ imply a fluctuation in the relative order of preference for the media object based on their CPC, which in turn leads to a behavior that is - at least partly - exploratory.